\def\bq{\begin{equation}}
\def\eq{\end{equation}}
\def\bqa{\begin{eqnarray}}
\def\eqa{\end{eqnarray}}
\def\bqb{\begin{eqnarray*}}
\def\eqb{\end{eqnarray*}}
\documentstyle[12pt]{article}\textwidth 16cm
\def\pr#1#2#3{ Phys. Rev. ${\bf{#1}}$ (#2) #3 }

\def\pl#1#2#3{ Phys. Lett. ${\bf{#1}}$ (#2) #3 }

\def\np#1#2#3{ Nucl. Phys. ${\bf{#1}}$ (#2) #3 }
\def\zp#1#2#3{ Z. Phys. ${\bf{#1}}$ (#2) #3 }

\def\L{ {\cal L }}
\def\O{ {\cal O }}

\global\nulldelimiterspace = 0pt

\def\frac#1#2{{{#1} \over {#2}}\,}  

\def\Dsl{\hbox{/\kern-.6700em\it D}} 
\def\dsl{\hbox{/\kern-.5300em$\partial$}}
\def\pxpsl{\hbox{/\kern-.5600em$p$}}
\def\ssl{\hbox{/\kern-.5300em$s$}}
\def\epssl{\hbox{/\kern-.5100em$\epsilon$}}
\def\delsl{\hbox{/\kern-.6300em$\nabla$}}
\def\lxpsl{\hbox{/\kern-.4300em$l$}}
\def\elxpsl{\hbox{/\kern-.4500em$\ell$}}
\def\kxpsl{\hbox{/\kern-.5100em$k$}}
\def\qxpsl{\hbox{/\kern-.5000em$q$}}
\def\sla#1{\raise.15ex\hbox{$/$}\kern-.57em #1}

\def\roughly#1{\mathrel{\raise.3ex
    \hbox{$#1$\kern-.75em\lower1ex\hbox{$\sim$}}}}
\def\lsim{\roughly<}
\def\gsim{\roughly>}

\begin{document}
\pagenumbering{arabic}
\thispagestyle{empty}

\begin{flushright} PM/93-26 \\ THES-TP 93/8 \\ September 1993 \end{flushright}
\vspace{2cm}
\begin{center}
{\Large\bf Vector boson Pair Production at Supercolliders; } \\
    {\Large\bf useful approximate helicity amplitudes}\dag{~}
 \vspace{1.5cm}  \\
 {\Large  G.J. Gounaris } \\
 Department of Theoritical Physics \\
University of Thessaloniki, Greece
\vspace {0.5cm}  \\
 {\Large  J. Layssac, F.M. Renard }\\
 Laboratoire de Physique
Math\'{e}matique \footnote {Unit\'{e} Associ\'{e}e au CNRS $n^0$
040768.}\\
Universit\'{e} de Montpellier II Sciences et
Techniques du Languedoc \\
Place E.Bataillon, Case 50 F-34095 Montpellier Cedex 5,
France
\vspace {2cm}  \\

 {\bf Abstract}
\end{center}
\noindent
We study vector boson pair production at $LHC$ and $SSC$,
taking into account the effects generated by the anomalous
vector boson and Higgs couplings induced by the
operators ${\cal O}_W$ and ${\cal O}_{UW}$, which are the only dim=6
 operators
preserving $SU(2)_c$. These operators lead to enhanced production
of transverse vector bosons, as opposed to the enhanced
production of longitudinal gauge bosons, induced in case
$M_H\gsim 1\ TeV$, by dim=4 terms already existing in the Standard Model
 lagrangian.
For vector boson pair masses larger than $1\ TeV$,
we establish very simple approximate  expressions for the
standard as well as the non-standard helicity amplitudes
for $q\bar q$ annihilation and vector boson fusion, which
accurately describe the physics. These expressions should
simplify the experimental search for such
interactions. We finally discuss the observability and the
disentangling of these interactions.
\vspace{1cm} \\
\dag{~}Work supported by the scientific cooperation program between CNRS
and EIE.
\setcounter{page}{0}

\clearpage

\section{Introduction}
   Studies of the gauge sector couplings constitute an important part of
the program of tests of  the
Standard Model and of the search for New Physics (NP) beyond it.
Such a study is particularly important if the new particles
connected to NP turn to be so heavy, that they cannot actually be
produced. Recently it has been
 recognized
that indirect tests [1,2] using high precision LEP 1 results at Z peak, are far
from having the required accuracy for a thorough study of NP,
and in any case they can only be meaningful for specific forms of the
new physics
models. Thus, direct measurements using vector boson pair
production have been shown to be essential for a comprehensive
study of the gauge boson (and Higgs) couplings induced by NP
[3-5]. Such direct tests will be first
achieved at LEP 2, where the study of $e^+e^- \to W^+W^-$ could
provide rough upper bounds of the
order of 10 percent, on the strength of the various three gauge
boson couplings [3]. Later on, a 500
 GeV
collider
could lower these bounds to
the order of 1 percent [4].  In between,
vector boson pair production in pp
collisions at LHC or SSC, should also be able to supply
information on the anomalous gauge boson and  Higgs
couplings, at the same level of accuracy  [6].\par
Departures from the standard Yang-Mills couplings (corrected by
Standard Model (SM) loop
effects) could arise from various sources like e.g.\ high order
effects involving  new heavy
particles, mixing with higher vector bosons, genuine departures from SM
structure, etc. Loop corrections are generally expected [7-9] to
lead to very small contributions to these departures,
marginally reaching the aforementioned 1 percent level.\par
However,
if the bosonic sector turns out to be strongly interacting, then
various processes may be enhanced [10]. This is a very  interesting
case, provided the bosonic sector is weakly coupled to fermions,
so that it avoids
the constraints from LEP 1 measurements. In a phenomenological analysis, a
 "natural" way
of selecting the NP operators that would not affect fermions may come from
a symmetry
principle. In a previous paper [6], we have
emphasized that the global  $SU(2)_c$
custodial symmetry provides a very interesting such principle,
which can be used to
guide our search for NP beyond SM. Such a symmetry is of course
not new. It has been invoked in the past in order to justify
the famous result $\rho \simeq 1$. It was also used later,
together with the implicit assumption
that at present energies NP only affects the purely scalar
sector [11]. On the basis of these assumptions it
was concluded that, in case $M_H \gsim 1TeV$,
NP induces a new strong
interaction among the \underline{longitudinal} W and Z gauge
bosons only [11].\par
In Ref.\ [6] we followed another root in exploring
the $SU(2)_c$ consequences, and allowed the $SU(2)_c$ invariant
field $\overrightarrow{W}_\mu$, to
participate in the NP induced operators.
We then concluded
that, provided NP is $SU(2)_c$ and CP symmetric and satisfies
$SU(2)\times U(1)$ gauge
invariance,  it can also induce the dim=6 operators
dubbed
${\cal O}_W$ and ${\cal O}_{UW}$
dimension. As we will see below, ${\cal O}_W$ involves couplings among the
gauge boson fields $W_{\mu \nu}$ only,
 whereas
$\O_{UW}$
describes an interaction containing the Higgs field also. In both cases,
 the
appearence of $W_{\mu \nu}$ guarantees that NP will
predominantly enhance the production of \underline{transverse} W
states at high energies [6], as opposed to the enhancement of the
longitudinal production considered before [11].
More explicitely, $\O_W$ and
$\O_{UW}$ will  always enhance
transverse-transverse (TT) pair production in $e^+e^-$ and $q\bar q$
 annihilation as well as in
boson-boson fusion.
Depending on the magnitude of the relevant couplings, these new
interactions may even induce a
new strong force among the transverse gauge bosons.
 We have found that the experimental search for such
interactions
is much easier than the study of the strong production of
longitudinal gauge bosons, arising in case the physical Higgs particle
is very heavy [6].
This is because in the $\O_W$ and
$\O_{UW}$ cases, it is sufficient to study  the total
rate of vector boson pair production, with no need for large background
subtraction.\par
According to Ref.\ [6] and the expected luminosities
 at LHC or SSC, the
SM rates of vector boson pair production should be
observable for invariant masses of the vector pair in the  range
M=1 to 2 TeV. In this range,  departures
from SM due to an
$\O_W$  anomalous coupling of the order of a
percent, should be easily observable at SSC/LHC [6].\par
In this paper we first show that for vector pairs in the TeV range,
there exist very simple approximate
 expressions for
all $q \bar q$ and vector fusion helicity amplitudes, corresponding
to both the standard and the non-standard interactions.
 These amplitudes are accurate at the percent level and
should be very useful for the
 experimental
analysis of the sensitivity to anomalous couplings. We then
study separately the  effects
of the operator $\O_W$ with a coupling $\lambda_W$, and $\O_{UW}$ with
 coupling $d$, on
$q\bar q$ annihilation and boson-boson fusion. \par
For studying the observability at SSC/LHC\@,
we sum up all processes for a given final state, and we ask for a 50 percent
 change in
the rate of the mass distribution $d\sigma/dM$ in the 1 to 2 TeV
range in order  to
conclude that we indeed have a signal of
New Physics (NP). A first study of the $ \O_W$ only has already been
done in [6]. Here we  add the
 Higgs
exchanges contributions to boson-boson fusion due to  $\O_{UW}$, and we
 study the
possibility to disentangle the effects of the two operators. We find that
disentangling is in principle possible, since different classes
of final states respond differently to $\lambda_W$ and  $d$.
We find for example that the ratios WZ/ZZ and $W\gamma/ZZ$
respond
 in opposite ways
as we increase  $|\lambda_W|$ and $|d|$. As a result, the
 discovery limits
appear to be of the order of 0.01 for $|\lambda_W|$, and 0.1 for $|d|$.\par
 The plan of the paper is the following. In Section 2 we recall the gauge
 invariant
forms of the $SU(2)_c$ preserving operators to be added to the Standard Model
lagrangian and give the corresponding 3-boson and 4-boson
vertices. The implied simple approximate
expressions for the $q\bar q$ annihilation
 and
boson-boson fusion amplitudes to two final gauge bosons are
given in the Appendices A and B\@. The distribution $d\sigma/dM$
is expressed using the parton model,
in terms of the quark and vector boson distribution functions
of the proton and the subprocess cross sections.
The results are presented in Section 3
for LHC or SSC energies, and invariant masses of the final
vector pair  in the
range of 1 to 2 TeV. The
observability of $\lambda_W$
 and $d$ effects is also discussed.
Finally, Section 4 summarizes the results.

\section{$SU(2)_c$ preserving couplings and subprocess \newline amplitudes}
It has been shown in [6] that, if we impose CP and $SU(2)_c$
symmetries as well as $SU(2)\times U(1)$ gauge
invariance and we restrict  to operators of dimension up to
six, then the effective lagrangian describing gauge
boson production at the various colliders is completely
described by \\ [.5cm]
\bq \L=\L_{SM}+\L_{NP}  \ \ \ \ \ , \eq

\bqa  \L_{SM}&=& -{1\over 2}\langle W_{\mu\nu}W^{\mu\nu}\rangle-{1\over4}
  B_{\mu\nu}B^{\mu\nu} +{v^2\over 4}\langle D_\mu
UD^\mu U^{\dagger}\rangle\nonumber\\[.5cm]
 \null & \null & -{v^2 M^2_H\over 8}
\left(\ {1\over 2}\ \langle UU^{\dagger}\rangle-1\right)^2+\
 \makebox{fermionic terms \ \ \ \ , } \eqa

\bq \L_{NP}=\lambda_W{g_2\over M^2_W}\O_W+d\O_{UW} \ \ \ \ , \eq

\noindent where

\bq \O_W={1\over3!}\left( \overrightarrow{W}^{\ \ \nu}_\mu\times
  \overrightarrow{W}^{\ \ \lambda}_\nu \right) \cdot
  \overrightarrow{W}^{\ \ \mu}_\lambda =-{2i\over3}
\langle W^{\nu\lambda}W_{\lambda\mu}W^\mu_{\ \ \nu}\rangle \ \ \
, \eq

\bq \O_{UW}=\langle (UU^{\dagger}-1)\ W^{\mu\nu} \
W_{\mu\nu}\rangle \ \ \ , \eq \\[.05cm]
and the standard Higgs doublet is given by \\
\bq U=\bigm(\widetilde \Phi\ \ , \ \Phi\bigm){\sqrt2\over v} \ \ \ \
, \eq

\bq \Phi=\left( \begin{array}{c}
      \phi^{\dagger}\\
{1\over\sqrt2}(v+H+i\phi^0) \end{array} \right) \ \ \ \ . \
\eq\\[.05cm]
\noindent Here the definitions $\langle A \rangle \equiv TrA$
and $\widetilde \Phi = i\tau_2 \Phi^* $ are  used.
Contrary to the heavy higgs scenario studied in [11-13], in the
present work we address
the case that the physical Higgs particle is rather light;
i.e.\@ $M_H \lsim 1TeV$. Thus, as explained in the Introduction, we
concentrate
on the possibility that the new $SU(2)_c$
invariant forces are realized via the transverse gauge boson
components. This can be seen explicitely
 from the
new three- and four boson vertices generated in the unitary
gauge by $\L_{NP}$ for $VW^+W^-$

\bq i\lambda_W{g_2\over M^2_W}\ W^3_{\nu\mu}
   \ W^{-\mu\lambda}\ W_{\ \ \lambda}^{+\ \ \nu}\ \ \ \ \ ,  \eq

for $VV^{\prime} W^+W^-$\\

\bqa \lambda_W{g^2_2\over M^2_W}  \bigm\{
         (W^-_\mu W^+_\nu- W^+_\mu W^-_\nu)\
       W^{-\nu\lambda}\ W_{\ \ \lambda}^{+\ \ \mu} \nonumber\\
   -W^3_{\mu\nu}W^3_\lambda\ (
W^{+\nu} W^{-\lambda\mu}+W^{-\nu} W^{+\lambda\mu})\nonumber\\
   + W^3_{\mu\nu}W^{3\nu}\ (
W^+_{\ \ \lambda} W^{-\lambda\mu}+W^-_{\ \ \lambda}\
W^{+\lambda\mu})  \bigm\}\ \ ,  \eqa

and for $HW^+W^-$\\

\bqa  d{g_2\over M_W}\ H\ \bigm\{ W^+_{\ \
\mu\nu}\ W^{-\mu\nu} \ \ \ \ \ \ \ \ \ \ \ \ \ \ \ \ \nonumber\\
 +{1\over2}(c^2_W\ Z_{\mu\nu}\ Z^{\mu\nu}
+s^2_W\ F_{\mu\nu}\ F^{\mu\nu}+2s_Wc_W\ F_{\mu\nu}\ Z^{\mu\nu})
\bigm\} \ , \eqa
where $g_2$ is the usual $SU(2)$ gauge coupling and
$W^3_\mu=c_WZ_\mu+s_WA_\mu$.\par

In pp collisions two types of subprocesses can
 produce
vector boson pairs, namely $q\bar q$ annihilation and boson-boson
fusion [6]. The $q\bar q$
annihilation to a pair of gauge bosons proceeds through standard
model quark exchange, and
through W,Z,$\gamma$
 formation
diagrams. Only the later ones are
sensitive to $\lambda_W$
contributions arising from three gauge boson couplings.
Although complete expressions for the sub-cross sections $d\sigma/dcos\theta$
 were
given in [6], it is instructive to look at the helicity amplitudes of this
 subprocess.

In the high energy approximation (which has been checked to be
valid at the percent level for vector pair masses M in the TeV
range), the $q\bar q \to VV^{\prime} $ amplitudes aquire the
very simple form given in Appendix A. From this we remark that at
high energies the SM
contribution to the amplitudes becomes energy-independent,
while the $\O_W$ contribution is proportional to
$\lambda_Ws/M^2_W$, in agreement with naive expectations
based on dimensional arguments and the equivalence theorem [12].
It can also be seen
that for the $q\bar q$ annihilation at high energies, there is never any
intereference between the
$\lambda_W$ and SM contributions, since
$\O_W$ contributes only  to amplitudes involving final
gauge bosons with equal transverse helicities, for which
the SM contributions vanish. This implies
that the cross section for the $q\bar q \to
VV^{\prime} $ subprocess
is always
quadratic in $\lambda_W$. Consequently,
the results for the $q\bar q$ annihilation to
any gauge boson pair presented in
Figs. 4 to 11 of [6] for $\lambda_W=0.01$, are also valid for
$\lambda_W=-0.01$. Thus, these
figures indeed provide a feeling of the general behaviour of a
non-vanishing $\O_W$ contribution for any sign of
$\lambda_W$.\par

We now turn to boson-boson fusion processes. These proceed through vector
boson exchanges and contact interactions, and in many cases  also
through Higgs boson exchanges. They contain therefore
$\lambda_W$ as well as $d$ contributions.
In [6], we numerically discussed only the $\lambda_W$
effects, using the exact tree level formulae of the SM
and $\O_W$ contributions. Explicit expressions for the
corresponding  vector fusion
amplitudes were not given though, because they were extremely lengthy.
However we have found that for vector pair masses M in the TeV range,
very simple approximate expressions
 for all the vector fusion
helicity amplitudes exist, which are accurate at the percent
level. They are given in Appendix B, and are very useful
for the phenomenological analysis of both
 standard
and non standard contributions.\par

It is instructive to contemplate on the  way
$\lambda_W$ and $d$ contribute to these amplitudes.
We first turn to the SM and $\O_W$ contributions and remark that
they appear in the channels  $\gamma W \to \gamma
W$, \  $\gamma W \to ZW$, \ $ZW \to
 \gamma W$, \
 $\gamma
\gamma \to WW$, \ $\gamma Z \to WW$, \ $ WW \to \gamma \gamma$,
\ $ WW \to \gamma Z$, \
 $ZZ \to WW$, \  $ WW \to ZZ$, \
$ZW \to ZW$ and $WW \to WW$ channels.  There exist
only three different
types of terms contributing to these channels; namely
purely SM terms that are energy-independent at high energies,
and terms proportional either to $\lambda_W
s/M^2_W$ or $(\lambda_W s/M^2_W)^2$, with their coefficients
depending only on the CM scattering angle $\theta$ for the sub-process. As in
the annihilation case, such a result is intuitively expected on the
basis of naive dimensional arguments and the equivalence theorem
[12]. According to the results in Appendix B, at most only two of these
types of terms can contribute in the same helicity amplitude,
creating rather small
interferences effects and a weak sensitivity of the
sub-process cross section on the sign of $\lambda_W$. This
sensitivity largely disappears though, when we consider
ratios of production cross sections for the various two gauge
boson channels; see below.\par

We next turn to the $\O_{UW}$ contribution determined by the
coupling $d$. Such a contribution exists for all vector fusion channels.
In particular it exists also for the
pure neutral channels $\gamma \gamma \to \gamma \gamma,\ \ \gamma Z \to \gamma
\gamma, \ \ ZZ \to \gamma \gamma, \ \ \gamma \gamma \to \gamma
Z, \ \ \gamma Z
 \to \gamma Z, \ \
\  ZZ \to
\gamma Z, \ \ \gamma \gamma \to ZZ, \ \gamma Z \to ZZ$, which
receive neither SM nor $\O_W$ contributions at tree level. In the
asymptotic formulae given in Appendix B, no assumption on the
Higgs mass was made. Nevertheless, if the Higgs mass is
comparable to $M_W$, these amplitudes acquire an even
simpler asymtotic form expressed in terms of two contributions
proportional to
$ds/M^2_W$ and $d^2 s/M^2_W$ respectively, with their
coefficients depending on
the scattering angle in the c.m.\ of the sub-process.

It also turns out that the $\O_{UW}$ contribution is invisibly
small unless $|d|\gsim 0.1$, for which the quadratic
$d^2$ term dominates the amplitudes, and any sensitivity of the production
cross sections on the sign of
$d$ is lost. Also worth noting from the amplitudes in Appendix B is the
fact, that this $d^2$ term contributes mainly to the production of
transverse gauge bosons.
Thus, provided $M_H \sim M_W$, the contribution of any of the
two $SU(2)_c$ induced operatos $\O_W$ and $\O_{UW}$ , is either
negligible or affects mainly the  transverse gauge
bosons.

\section{Observability at LHC or SSC}
We have computed $d\sigma/dM$ for pp collisions as explained in
[6]. Subprocess  cross
sections are convoluted with quark or vector boson distribution
functions inside  the
proton. The same cuts on the rapidity Y and the photon transverse momentum are
 applied; i.e.\ $|Y|\leq2$, $|p_{\gamma T}|\leq0.05TeV$.
For a given final vector boson pair we sum all processes with different initial
 states.
In particular we note the importance of including initial states containing
 photons as
now transverse states are dominant.
As mentioned in Section 1 there is now essentially no background
problem as the  signal
we are looking for affects all states and predominantly the
leading TT ones.\par

Consequently, the experimental search for the $\O_W$ and $\O_{UW}$ effects
discussed here, is much simpler than the search
for the strong interactions affecting the
longitudinal gauge bosons discussed in [11,13,14]. In this later
case the interesting signal is completely contained in the LL
final states, whose study requires the subtraction of a large SM
background due to the transverse gauge boson productions in
$q\bar q$ annihilation and vector fusion sub-processes.

One could also worry about a possible background from gluon-gluon fusion
processes that can contribute, through a quark loop, to the SM
result for $W^+W^-$, $ZZ$, $Z\gamma $ and $\gamma \gamma $ production.
However, the results of [15] show that for $M_H  \sim M_W $,
$m_t=120GeV$, and $M_{ZZ}\gsim 1 TeV$ we have
\bq
\frac{d\sigma (gg \to ZZ)/dM_{ZZ}}{d\sigma (q\bar q \to ZZ)/dM_{ZZ}}
\lsim {0.25} \mbox{ for SSC \ \ \ . \  }
 \eq\\[.01cm]
For LHC the r.h.s. of the inequality (11) becomes $\sim 0.1$,
and in both (SSC and LHC) cases it slowly decreases as $M_{ZZ}$
increases. These results are understandable on the basis of the
softness of the gluon distribution functions, and allow us to
expect that the gluon fusion background should not be very
important for $M\gsim 1TeV$. In any case, we have neglected this
SM background in the present work.\par

 There are certainly uncertainties introduced by the quark and vector boson
distribution functions inside the proton. For this reason we define the
 discovery
limit for an $\O_W$ or $\O_{UW}$ interaction, by asking for a 50
percent change in the rate as compared to the SM
prediction. This is what gives the order of magnitude limits of 0.01 on
 $|\lambda_W|$ and 0.1
for $|d|$. We also stress the fact that \underline{ratios} of
production rates to
different final states, should be less sensitive to the
uncertainties in distribution
functions.\par

We also emphasize that in the present case, where all
helicities of the final (and initial) gauge bosons are summed
over, the $q\bar q$ annihilation and vector fusion mechanisms
are both important,
for producing gauge boson pairs in the range $1 TeV \lsim M \lsim
2 TeV$.  This
result was first noticed in [6] and it contrasts sharply with
the situation in
Refs.[13,14], where the interest is focused to the production of
longitudinal gauge bosons at lower M values and vector fusion is
ignored.\par

We now turn to the comparison of $\O_W$ and $\O_{UW}$ effects.
We  classify the
various contributing channels into three classes:\par
a) Channels in which the enhancements for the limiting values
 $|\lambda_W|=0.01$ and $|d|=0.1$
 are similar, like e.g.\@
$WZ \to WZ, W \gamma \to W \gamma$. Fig.1 illustrates the case
of $W^+Z \to W^+Z$.\par

  b) Channels in which the $|d|=0.1$ effects are larger
than $|\lambda_W|=0.01$ ones. This happens when the standard contribution
is depressed or absent, like e.g.\@ in the purely neutral channels as
 $ZZ \to ZZ$, $\gamma \gamma \to \gamma
 \gamma$,
$\gamma Z \to \gamma Z$, where $\lambda_W$ does
not contribute.  Fig.2 illustrates the ZZ case.\par

c) Finally we consider channels in which $d$ effects are weaker
than $\lambda_W$ ones. This is the case of the WW channels, like $\gamma \gamma
 \to WW$, $WW \to WW$,
 because of the presence of large photon contributions and strong charged
 W couplings.\par

 Fig.3 and 4 show the results of summing all initial states for given final
states.
As expected from the amplitudes given in Appendix A and B,
changing the sign of  $\lambda_W$ and
 $d$ leads to  similar effects.
So roughly no ambiguity should arise  from these
 signs, in
the 1 to 2 TeV range of M.\par
With the expected LHC or SSC luminosities and the need to
observe at least 10
 events in order to establish NP,
the figures confirm the discovery limits that we have announced.
The disentangling of $\lambda_W$ and $d$ can be done by
comparing departures to  Standard
predictions in the three above
 classes of channels. A spectacular way of illustrating
 these
properties consists in plotting specific ratios, like those
presented in Figs. 5 and 6.
 As we see there, irrespective of
signs a remarkable separation of the $\lambda_W$ and $d$
effects is
 provided by the
ratios WZ/ZZ and $W\gamma/ZZ$, which can further be tested by
also studying
ratios like $WZ/\gamma\gamma$, $WZ/W\gamma$.\par

\section{Final discussion}
 We have examined the observable consequences of the existence
of New Physics in  the
gauge boson sector that   preserves $SU(2)_c$ global symmetry in addition
 to the $SU(2)\times U(1)$
 spontanously broken gauge
invariance. This is a natural way to allow for non
standard self-boson couplings that avoid strong constraints from high precision
 tests
at LEP 1. Two effective interactions $\lambda_W\O_W$ and
$d\O_{UW}$ modify vector boson
self-couplings and Higgs-vector boson couplings respectively. They lead to TT
dominance at high energies, a complementary case to the one
generated by the $SU(2)_c$ conserving
 scalar
sector leading to LL dominance. In the  TT case there is no
background problem contrarily
 to the
LL case. We have then discussed the search of such interactions
at LHC or SSC.\par
The first main result of this paper is the establishment of very
simple and useful
 high
energy approximate expressions for the helicity amplitudes of all subprocesses
 with
$q\bar q$ annihilation and boson-boson fusion. These expressions
 allow us to understand the
 features of
$\lambda_W$ and $d$ effects and to compute them
very easily, with an accuracy of 1 percent for $1TeV \lsim M
\lsim 2TeV$. They should be very useful for phenomenological
analyses in pp and
 $e^+e^-$
collisions.\par
 Our second main result concerns the disentangling of
 $\lambda_W$ and $d$
effects. Summing over $q\bar q$ and
 boson-boson
initial states we have found three classes of final states
characterized by their M dependence for various
 $\lambda_W$ and
$d$. The most spectacular such behaviour is observed in the
ratios WZ/ZZ and
 $W\gamma/ZZ$
 shown
in Fig.6. As we see there, $\lambda_W$ and $d$ give opposite
departures to the Standard Model
 predictions,
irrespective of the signs of the couplings. These ratios can
therefore be used in order to disentangle $\lambda_W$ and $d$
contributions. Other ratios can  be used in order to further test
whether this disentangling is correct.\par
 The
discovery limit of\@ 0.01 for $\lambda_W$ and 0.1 for $d$ have different
 implications. In the
pure gauge sector, such $\lambda_W$ measurement at SSC or LHC
should improve by an order of
 magnitude the anticipated  LEP2
 result, and by two orders of magnitude the result from LEP1\@.
Thus the SSC/LHC accuracy is comparable to the
 one
expected from NLC. However the new feature here is the
possibility to study
 many
different channels, and to have access to the other $SU(2)_c$ conserving
 operator
$\O_{UW}$, which affects Higgs couplings not directly accessible in
 $e^+e^-$.
Concerning the $\O_{UW}$ effects, we should also remark that
although the accuracy on $d$ is weaker than the one on
$\lambda_W$, it will constitute
 an important
insight to the Higgs sector.
Finally, the complementarity to the $SU(2)_c$ LL interactions,
 should
allow to get a view of the structure of the interactions associated to the
 mechanism of
mass generation, that is still a most puzzling domain in particle
 physics.\vspace{0.5cm}\\
{\Large \bf Acknowledgements}\vspace{0.5cm}\par
 One of us (F.M.R.) wishes to thank the Department of Theoretical Physics of
the
 University
of Thessaloniki for the warm hospitality and the kind help that he received
 during his
stay and the
preparation of this paper.
\newpage

\def\sw{s^2_W}
\def\cw{c^2_W}
\def\tt{\tau_3}
\def\cot{\cos\theta}
\def\sit{\sin\theta}
\def\ed{e^2}
\def\mw{M_W^2}
\def\lw{\lambda_W}
\def\lsm{\biggm( {\lw s\over\mw}\biggm) }
\def\rd{\sqrt2}

\def\ep#1#2{(\epsilon_{#1}\epsilon_{#2})}
\def\dh#1{ {1\over D_H({#1})} }
\def\nh#1{  D_H({#1}) }
\def\co{\biggm[}
\def\cf{\biggm]}

\renewcommand{\theequation}{A.\arabic{equation}}
\setcounter{equation}{0}

{\Large\bf Appendix A : Helicity amplitudes for $q\bar q \to
VV'$ \newline processes at High Energy}\\

Neglecting quark
masses, the quark annihilation amplitudes vanish, unless
 $q$ and $\bar q$
have opposite helicities denoteted by $\lambda $ and $-\lambda$
respectively. The helicities of the vector states $V$ and $V'$
are denoted by $\tau $ and $\tau'$ respectively. The
 helicity amplitudes for the various annihilation processes
$q(\lambda)\bar q(-\lambda) \to V(\tau)V'(\tau')$ are described
by $F^{L}_{ \tau \tau'}$ for $\lambda =1/2$, and by $F^{R}_{ \tau \tau'}$
for $\lambda =-1/2$. The Jacob-Wick phase conventions [16] are
used, and the antifermion
wave function is such that
$u(k,\pm)=C \bar v^{\tau}(k,\pm)$; i.e.\@ charge conjugation implies
$e_{L}^-\ (e^-_{R})\to e_{L}^+\ (e^+_{R})$ exactely.
Imposing CP invariance, there are at most 6
independent $F^{L}_{ \tau \tau'}$ helicity amplitudes,
and another 6 $F^{R}_{ \tau \tau'}$ ones.  The
normalization is such that the  differential cross section
 is given by  \par
\bq  {d\sigma\over dcos\theta} = C \sum_{\lambda \tau \tau'}|F_{\lambda
\tau \tau'}|^2
\eq
where the coefficient\par
\bq C = {1\over128N_c\pi s}\, {2p_V\over\sqrt{s}} \eq
 takes care of the average over the initial $q \bar q$ spin states and colour
factor $N_c$. Here and
below the usual Mandelstam variables for the sub-processes are
denoted by $s,t,u$, while $p_V$ is the cm momentum of the final bosons
and $\theta $ is the angle between $W^-$ and the fermion.
Q is the fermion
charge in units of e, and
$\tau_3=-1$ if the initial quark is $d$, and $\tau_3=+1$ if it
is  $u$. The amplitudes for the various channels for $s \gsim 1
TeV^2 $ are:
\vspace{1cm}\par

\fbox{ {\bf $e^+e^-,\ d\bar d,\ u\bar u\to W^+ W^-$ }}\par

\bqa
F^L_{++} &=& F^L_{--}=\:\tt\: {\ed\over4\sw}\lsm\sit\nonumber\\[.1cm]
F^L_{+-} &=& -{\ed\over2\sw}\sit\:{(1-\tt\cot)\over1+\cot}\ \ \ ;\ \
F^L_{-+} = {\ed\over2\sw}\sit\:{(1-\tt\cot)\over1-\cot} \nonumber\\[.1cm]
F^L_{LL}&=& \tt{\ed\over2\cw}\sit\:(|Q|-1+{1\over2\sw})\nonumber\\[.1cm]
F^R_{LL}&=& Q{\ed\over2\cw}\sit \eqa

\null\newpage
 \fbox{ {\bf $d\bar u\to  W^- Z$ }}\par
\bqa
F^L_{++} &=& F^L_{--}=-\:{\ed \over2\rd}\: {c_W\over
s^2_W}\:\lsm\sit\nonumber\\
F^L_{+-} &=& -\:{\ed\over\rd
c_W\sw}\:\:{\sit\over1+\cot}\ \bigm(\cw\cot-{\sw\over3}\bigm)\nonumber\\[.1cm]
F^L_{-+}& =&{\ed\over\rd c_W\sw}\:\: {\sit\over1-\cot}\
\bigm(\cw\cot-{\sw\over3}\bigm)\nonumber\\[.1cm]
F^L_{LL}&=& -\:{\ed\over2\rd\, \sw}\:\sit\eqa
\null\vspace{1cm}\par

 \fbox{ {\bf $d\bar u\to  W^- \gamma$ }}\par
\bqa
F^L_{++} &=& F^L_{--}=\:-\:{\ed \over2\rd\, s_W}\:\sit\lsm\nonumber\\[.1cm]
F^L_{+-} &=& -\:{\ed\over\rd\, s_W}\ \
 {\sit\over1+\cot}\ \bigm(\cot+{1\over3}\bigm)\nonumber\\[.1cm]
F^L_{-+} &=& {\ed\over\rd\, s_W}\ \
 {\sit\over1-\cot}\ \bigm(\cot+{1\over3}\bigm)\eqa
\null\vspace{1cm}\par

 \fbox{ {\bf $e^+e^-,\ d\bar d,\ u\bar u\to  \gamma \gamma$ }}\par
\bqa
F^L_{+-} &=& F^R_{-+}=\:2\ed Q^2\ {(\cot-1)\over\sit }\nonumber\\[.1cm]
F^L_{-+} &=& F^R_{+-}=\:2\ed Q^2\ {(\cot+1)\over\sit }\eqa
\null\vspace{1cm}\par

 \fbox{ {\bf $e^+e^-,\ d\bar d,\ u\bar u\to  Z\gamma$ }}\par
\bqa
F^L_{+-} &=& {\ed Q\over s_W c_W}\:(\tt-2Q\sw)\ {(\cot-1)\over\sit}
\nonumber\\[.1cm]
F^L_{-+} &=& {\ed Q\over s_W c_W}\:(\tt-2Q\sw)\ {(\cot+1)\over\sit}
\nonumber\\[.1cm]
F^R_{-+} &=&\:{2\ed Q^2s_W\over c_W}\ {(\cot-1)\over\sit}\nonumber\\[.1cm]
F^R_{+-} &=&\:{2\ed Q^2s_W\over c_W}\ {(\cot+1)\over\sit}\eqa
\null\vspace{1cm}\par

 \fbox{ {\bf $e^+e^-,\ d\bar d,\ u\bar u\to  ZZ$ }}\par
\bqa
F^L_{+-} &=& {\ed \over2\sw\cw}\:(\tt-2Q\sw)^2\ {(\cot-1)\over\sit}
\nonumber\\[.1cm]
F^L_{-+} &=& {\ed \over2\sw\cw}\:(\tt-2Q\sw)^2\ {(\cot+1)\over\sit}
\nonumber\\[.1cm]
F^R_{-+} &=&-\:{2\ed Q^2\sw\over \cw}\ {(\cot-1)\over\sit}\nonumber\\[.1cm]
F^R_{+-} &=&-\:{2\ed Q^2\sw\over \cw}\ {(\cot+1)\over\sit}\eqa
\null\vspace{1cm}\par

According to (A3)-(A8), $\O_W$
contributions at high energies exist only for the amplitudes
$F_{++}$ and $F_{--}$, where
both final gauge bosons have equal and transverse heleicities.
These $\O_W$ contributions behave like $\lw s/\mw$ to leading
order. Since these amplitudes receive no standard contribution,
there exists no appreciable
interference between standard and $\lw$ terms at high energies. This explains
the lack of sensitivity to the sign of $\lw$ in these processes.
\newpage

\renewcommand{\theequation}{B.\arabic{equation}}
\setcounter{equation}{0}
\setcounter{section}{0}

{\Large\bf Appendix B : Helicity amplitudes for boson-boson
fusion processes at  High
Energy}\\

 In general there are 81  helicity amplitudes $F_{\lambda \lambda' \mu \mu'}$
for each vector boson-vector boson fusion
process  $V_1(\lambda) V_2(\lambda') \to
V_3(\mu) V_4(\mu')$. Taking into account parity conservation,
(which is valid at tree
level for the self-boson interactions contained in SM and $\O_W$,
 $\O_{UW}$), implies the relation
\bq
F_{\lambda\lambda'\mu\mu'}(\theta)~=~F_{-\lambda-\lambda'-\mu-\mu'}(\theta)
\ \ \  , \ \ \ \ \eq
\noindent
which reduces the number of independent amplitudes to 41. In
specific channels this
 number is
further reduced due e.g.\@ to the absence of helicity zero states for photons,
 the symmetrization
for identical particles, charge conjugation relations, etc. Here and
below $\theta$ is the angle between $V_1$ and $V_3$. The
normalization of these amplitudes is defined by noting that the
differential cross section in c.m.\@ is given by\par

\bq  {d\sigma(\lambda \lambda' \mu \mu')\over dcos\theta} = C |F_{\lambda
\lambda'
 \mu
 \mu'}|^2 \ \ \ \ , \ \ \ \ \eq
 where the coefficient\par
\bq  C = {1\over32\pi s}\,{p_{34}\over p_{12}} \eq
includes \underline{no spin average}.  This later choice is
motivated by the fact that, inside the proton, different vector
boson distribution functions  occur for different initial
 helicity
states. Finally  $p_{12}$, $p_{34}$ in (B.3)
denote the cm momenta of the initial and final boson pairs
respectively.\par

\section{Standard and $\O_W $ contributions}
 In this part we give the SM and $\O_W$ contributions to the amplitudes arising
from exchanges of gauge bosons, as well from exchanges of
a Higgs boson with  $M_H=M_W$.
The corrections to be added to these expressions, due to the fact
that in general $M_H
\neq M_W $, are given in the second
part of this Appendix together with the $ \O_{UW}$
contributions. The motivation for this choice is that it leads to SM
contibutions which are always finite, even in the presence of loops.
We thus have for $s \gsim 1 TeV^2 $ : \vspace{1cm}\par

\fbox{ {\bf $ \gamma W\to \gamma W$ }}\par
\bqa
 F_{++++}&=&F_{----}=\:-\:\ed\biggm\{ {4\over1+\cot}\:+\:\lsm^2\
{\cot\over4}\biggm\} \nonumber\\[.1cm]
F_{+++-}&=&F_{++-+}=\ F_{+-++}=\ F_{-+++}=  \nonumber\\
F_{---+}&=&F_{--+-}=\ F_{-+--}=\ F_{+---}= \ed\ {(1-\cot)\over2}\ \lsm
\nonumber\\
F_{+--+}&=&F_{-++-}=\:-\:\ed\ {(1-\cot)^2\over1+\cot}\nonumber\\[.1cm]
F_{+-+-}&=&F_{-+-+}=\:-\:\ed\ (1+\cot)\biggm\{1+{3-\cot\over16}\
\lsm^2\biggm\}\nonumber\\[.1cm]
F_{++--}&=&F_{--++}=\ed\ \lsm
\biggm\{1-\cot-{(3+6\cot-\cos^2\theta)\over16}\ \lsm\biggm\}\nonumber\\[.1cm]
F_{+L+L}&=&F_{-L-L}=-2\ed  \eqa
\null\vspace{1cm}\par

\fbox{ {\bf $ \gamma W\to Z W$ }}\vspace{.5cm}\par
The purely transverse amplitudes are identical with those in
$\gamma W\to \gamma W$,
provided we replace $\ed\to \ed c_W/s_W $. The
amplitudes involving longitudinal bosons are
 \bqa
F_{++LL}&=&F_{--LL}=\ {\ed\over4s_W}\cot\ \lsm\nonumber\\
F_{+-LL}&=&F_{-+LL}=\ -\ {\ed\over2s_W}\ (1-\cot) \nonumber\\
F_{+LL-}&=&F_{-LL+}=\ {\ed\over8s_W}\ (\cot-3)\ \lsm\nonumber\\
F_{+LL+}&=&F_{-LL-}=\ -\ {\ed\over s_W}\ {(\cot-1)\over\cot+1} \eqa
 \null\vspace{1cm}\par

\fbox{ {\bf $ \gamma\gamma\to W W$ }}
\bqa
F_{++++}&=&F_{----}=\ {8\ed\over\sin^2\theta}\nonumber\\[.1cm]
F_{+++-}&=&F_{++-+}=\ F_{+-++}=\ F_{-+++}=  \nonumber\\
F_{---+}&=&F_{--+-}=\ F_{-+--}=\ F_{+---}=\ -\ed\,\lsm\nonumber\\[.1cm]
F_{+--+}&=&F_{-++-}=\ed(1-\cot)\biggm\{{2\over1+\cot}+{3+\cot\over16}
\lsm^2\biggm\}\nonumber\\[.1cm]
F_{++--}&=&F_{--++}=\ed\lsm
  \biggm\{-2\ +\ {3-\cos^2\theta\over8}\lsm\biggm\}\nonumber\\[.1cm]
F_{+-+-}&=&F_{-+-+}=\ -\ed\ (1+\cot)
\biggm\{{2\over\cot-1}+{(\cot-3)\over16}\ \lsm^2\biggm\}\nonumber\\[.1cm]
F_{+-LL}&=&F_{-+LL}=2\ed  \eqa
\null\vspace{1cm}\par

\fbox{ {\bf $ \gamma Z\to W W$ }} \vspace{.5cm}\par
The purely transverse amplitudes are identical with those in $\gamma\gamma\to
WW$ provided we replace $\ed\to\ed c_W/s_W$.
The amplitudes involving longitudinal bosons are
\bqa
F_{+LL+}&=&F_{-LL-}=\ -\ {\ed\over s_W}\ {2\over1+\cot} \nonumber\\[.1cm]
F_{+L-L}&=&F_{-L+L}=\ -\ {\ed\over 8s_W}\ (3+\cot)\ \lsm \nonumber\\[.1cm]
F_{+L+L}&=&F_{-L-L}=\ {\ed\over s_W}\ {2\over1-\cot} \nonumber\\[.1cm]
F_{+LL-}&=&F_{-LL+}=\ {\ed\over 8s_W}\ (3-\cot)\ \lsm \eqa
\null\vspace{1cm}\par

\fbox{ {\bf $ Z Z\to W W$ }}\vspace{.5cm}\par
The purely transverse amplitudes are identical with those
in $\gamma\gamma\to WW$
provided we replace $\ed\to \ed\cw/\sw $. The longitudinal
amplitudes are given by
\bqa
F_{L++L}&=&F_{L--L}=F_{+LL+}=F_{-LL-}=-\,{\ed\over\sw}\
{(4\cw-1-\cot)\over2c_W(1+\cot)}\nonumber\\[.1cm]
F_{+L-L}&=&F_{-L+L}=F_{L-L+}=F_{L+L-}= -\,{\ed c_W\over\sw}\
{(3+\cot)\over8}\ \lsm\nonumber\\[.1cm]
F_{L+L+}&=&F_{L-L-}=F_{+L+L}=F_{-L-L}=-\,{\ed\over\sw}\
{(1-\cot-4\cw)\over2c_W(1-\cot)}\nonumber\\[.1cm]
F_{+LL-}&=&F_{-LL+}=F_{L-+L}=F_{L+-L}={\ed c_W\over\sw}\
{(3-\cot)\over8}\ \lsm\nonumber\\[.1cm]
F_{+-LL}&=&F_{-+LL}={\ed\over2}\,\left({c_W\over s_W}\,-\,
{s_W\over c_W}\right)^2\nonumber\\[.1cm]
F_{LL+-}&=&F_{LL-+}={\ed\over2\sw}\ \ \ ;\ \ \ F_{LLLL}={\ed\over\sw}\
{(5+3 cos^2\theta)\over4\sin^2\theta}  \eqa
\null\vspace{1cm}\par
\fbox{ {\bf $ZW \to \gamma W$, $ WW\to \gamma\gamma$, $WW\to \gamma Z$ and $
WW\to ZZ$ }}\vspace{.5cm}\par
These amplitudes are respectively identical with those in $\gamma W \to ZW$,
$\gamma\gamma\to WW $, $\gamma
Z\to WW$ and $ZZ\to WW$ by interchanging helicities between the
initial and final state.
\null\vspace{1cm}\par

\fbox{ {\bf $ Z W\to Z W$ }}\vspace{.5cm}\par
The purely transverse amplitudes are identical with those in $\gamma W\to
\gamma W $
provided we replace $\ed\to \ed \cw/\sw $. The longitudinal
amplitudes are
\bqa
F_{LL\pm\pm}&=&F_{\pm\pm LL}= {\ed c_W\over4\sw}\ \cot\ \lsm \nonumber\\[.1cm]
F_{LL\pm\mp}&=&F_{\pm\mp LL}= {\ed \over2c_W}\ (1+{\cw\over\sw}\cot)
\nonumber\\
F_{\pm LL\mp}&=&F_{L\pm\mp L}={\ed c_W\over8\sw}\ (\cot-3)\ \lsm
\nonumber\\[.1cm]
F_{L\pm\pm L}&=&F_{\pm LL\pm}={\ed
\over 2c_W\sw}\biggm\{ 1+2\cw\ {1-\cot\over1+\cot}\biggm\}\nonumber\\[.1cm]
F_{\pm L\pm L}&=&-\, {\ed\over2\cw\sw}\ (\cw-\sw)^2\ \ \ \ ;\ \ \
  F_{L\pm L\pm}=\ -\,{\ed\over2\sw}\nonumber\\[.1cm]
F_{LLLL}&=&-\,{\ed\over\sw}\ {(4+\cot+\cos^2\theta)\over4(1+\cot)}  \eqa
\null\vspace{1cm}\par

\fbox{ {\bf $ W^+ W^-\to  W^+W^-$ }}
\bqa
F_{\pm\pm\pm\pm}&=&{\ed\over\sw}\biggm\{ {4\over1-\cot}\,-\,{\cot\over4}\,
\lsm^2\biggm\} \nonumber\\[.1cm]
F_{\pm\mp\mp\mp}&=&F_{\mp\pm\mp\mp}=F_{\mp\mp\pm\mp}=F_{\mp\mp\mp\pm}=
-\,{\ed\over2\sw}(1+\cot)\, \lsm \nonumber\\[.1cm]
F_{\pm\mp\mp\pm}&=&{\ed\over\sw}\,(1-\cot)\biggm\{1+{3+\cot\over16}\lsm^2
\biggm\}\nonumber\\[.1cm]
F_{\pm\pm\mp\mp}&=&{\ed\over\sw}\, \lsm\biggm\{
-1-\cot\,+{3-6\cot-\cos^2\theta\over16}\, \lsm \biggm\}  \nonumber\\[.1cm]
F_{\pm\mp\pm\mp}&=& {\ed\over\sw}\ {(1+\cot)^2\over1-\cot}  \nonumber\\[.1cm]
F_{LL\pm\mp}&=&F_{\pm\mp LL}={\ed\over2\sw}\,(1+\cot)\nonumber\\[.1cm]
F_{L\pm L\pm}&=&F_{\pm L\pm L}={\ed\over\sw}\
{(1+\cot)\over(1-\cot)}\nonumber\\
F_{L\pm L\mp}&=&F_{\pm L\mp L}=-\,{\ed\over8\sw}\,(3+\cot)\, \lsm
\nonumber\\[.1cm]
F_{LL\pm\pm}&=&F_{\pm\pm L L}={\ed\over4\sw}\,\cot\, \lsm \nonumber\\[.1cm]
F_{LLLL}&=&-\,{\ed\over4\sw\cw}\biggm\{
2\cw+{3+\cos^2\theta\over\cot-1}\biggm\}   \eqa
\null\vspace{1cm}\par

\fbox{ {\bf $ W^+ W^+\to  W^+W^+$ }}
\bqa
F_{\pm\pm\pm\pm}&=&-\,{\ed\over\sw}\ {8\over\sin^2\theta}\nonumber\\[.1cm]
F_{\mp\pm\pm\pm}&=&F_{\pm\mp\pm\pm}=F_{\pm\pm\mp\pm}=F_{\pm\pm\pm\mp}
  ={\ed\over\sw}\, \lsm \nonumber\\[.1cm]
F_{\pm\mp\mp\pm}&=&-\,{\ed\over\sw}\ (1-\cot)\biggm\{
{2\over1+\cot}+{3+\cot\over16}\ \lsm^2 \biggm\}  \nonumber\\[.1cm]
F_{\pm\pm\mp\mp}&=& {\ed\over\sw}\,
\lsm\biggm\{2\ +\, {\cos^2\theta-3\over16}\, \lsm \biggm\} \nonumber\\[.1cm]
F_{\pm\mp\pm\mp}&=&-\,{\ed\over\sw}\ (1+\cot)\biggm\{{2\over1-\cot}\,+\,{3-\cot
\over16}\ \lsm^2 \biggm\} \nonumber\\[.1cm]
F_{L\pm L\pm}&=&F_{\pm L\pm L}=\ -\,{2\ed\over\sw(1-\cot)} \nonumber\\[.1cm]
F_{\pm LL\pm}&=&F_{L\pm \pm L}=\ {2\ed\over\sw(1+\cot)} \nonumber\\[.1cm]
F_{\pm LL\mp}&=&F_{L\pm \mp L}={\ed\over\sw}\ {(\cot-3)\over8}\, \lsm
\nonumber\\
F_{L\pm L\mp}&=&F_{\pm L\mp L}={\ed\over\sw}\ {(\cot+3)\over8}\, \lsm
\nonumber\\
F_{LLLL}&=&{\ed\over2\cw}\biggm\{1-{4\over\sw\sin^2\theta} \biggm\}
\eqa
\null\vspace{1cm}\par

\fbox{ {\bf $ ZZ\to ZZ$ }}
\bqa
F_{LLLL}&=&-\,{3\ed\over4\sw}\nonumber\\
F_{LL\pm\mp}&=&F_{\pm\mp LL}=F_{L\pm\pm L}=F_{\pm LL\pm}=-F_{L\pm
L\pm}=-F_{\pm
L\pm L}={\ed\over2\sw\cw }
\eqa
\vspace{1cm}
\section{ $ \O_{UW}$ contributions, and corrections to the
standard model predictions due to $M_H\neq M_W$}
It is convenient to express these contributions to the helicity
amplitudes as
functions of the intial and final helicities. The vector boson  helicities
are indicated in parentheses below, where  ($z=cos\theta$)
the vector fusion process is writen as
\bq V_1(\lambda)\  V_2(\lambda') \to V_3(\mu)\  V_4(\mu') \ \ \ . \ \ \eq
The masses of the vector bosons are denoted by $m_j$  for
$(j=1,...,4)$, while $\epsilon_1$, $\epsilon_2$ denote the
polarization vectors for the intials boson states, and
$\epsilon_3$, $\epsilon_4$ the complex conjugate ones for the
final states.
We then use the following definitions at high energies :\par

\bqa
\ep{1}{2} &=&-\,{s\over2 m_1 m_2}\ (1-\lambda^2)(1-\lambda'^2) \nonumber\\
\ep{1}{3} &=&{s(1-\cot)\over4 m_1 m_3}\ (1-\mu^2)(1-\lambda^2) \nonumber\\
\ep{1}{4} &=&-\,{s(1+\cot)\over4 m_1 m_4}\ (1-\mu'^2)(1-\lambda^2) \nonumber\\
\ep{2}{3} &=&-\,{s(1+\cot)\over4 m_2 m_3}\ (1-\mu^2)(1-\lambda'^2) \nonumber\\
\ep{2}{4} &=&{s(1-\cot)\over4 m_2 m_4}\ (1-\mu'^2)(1-\lambda'^2) \nonumber\\
\ep{3}{4} &=&-\,{s\over2 m_3 m_4}\ (1-\mu^2)(1-\mu'^2) \eqa
\bqa
V_{12}&=&{s\over4}\, \lambda^2\lambda'^2(1+\lambda\lambda')\ \ \ \ ;\ \ \
 V_{34}={s\over4}\, \mu^2\mu'^2(1+\mu\mu')\nonumber\\
V_{13}&=&{s\over8}\,(1-\lambda\mu)(1-\cot)\mu^2\lambda^2\ \ \ \ ;\ \ \
V_{24}={s\over8}\,(1-\lambda'\mu')(1-\cot)\mu'^2\lambda'^2 \nonumber\\
V_{14}&=&{s\over8}\,(1-\lambda\mu')(1+\cot)\mu'^2\lambda^2\ \ \ \ ;\ \ \
V_{23}={s\over8}\,(1-\lambda'\mu)(1+\cot)\mu^2\lambda'^2
\eqa
\bq Z_{ij}=\ep{i}{j}\ {M_W\over\cw} - {2d\over M_W}\cw\ V_{ij} \eq
The Higgs propagator is written as
\bq D_H(x)\equiv x-{\cal M}^2_H  \eq
where ${\cal M}_H\equiv M^2_H$ for $x=t$ or $u$, and ${\cal
M}^2_H\equiv M^2_H-i\ M_H\Gamma_H$ when $x=s$. \par
Using these definitions, we describe by $F_H$ the sum of the $\O_{UW}$
contributions, and the corrections implied by
the SM higgs exhange interactions in case
$M_H \neq M_W$. We find

\bq  F_H(\gamma W\to\gamma W)\ =\ {2g^2_2\sw d\over M_W}\
V_{13}\biggm[\ep{2}{4}
M_W-{2d\over M_W} V_{24} \biggm] \dh{t} \eq

\bq F_H(\gamma W\to ZW)\ =\ {c_w\over s_W}\, F_H(\gamma W\to\gamma W) \eq
\bq F_H(Z W\to \gamma W)\ =\ F_H(\gamma W\to Z W) \eq

\bqa & & F_H(ZW\to ZW)={g^2_2\mw\over\cw}\,\ep{1}{3}\ep{2}{4}\co
 {1\over t-\mw}-\dh{t}\cf+  \nonumber\\
  & &\null +2g^2_2d\co {1\over\cw}\,\ep{1}{3} V_{24}
+\cw\ep{2}{4}V_{13}-{2d\over\mw}\cw V_{13}V_{24}\cf\dh{t} \eqa

\bq F_H(\gamma \gamma\to WW)={2g^2_2\sw d\over
M_W}\,V_{12}\co\ep{3}{4}M_W-{2d\over M_W}V_{34}\cf\dh{s}  \eq

\bq F_H(\gamma Z\to WW)={c_W\over s_W} F_H(\gamma\gamma\to WW) \eq

\bqa & & F_H(ZZ\to WW)={g^2_2\mw\over\cw}\,\ep{1}{2}\ep{3}{4}\co
 {1\over s-\mw}-\dh{s}\cf+  \nonumber\\
  & &\null +2g^2_2d\co {1\over\cw}\,\ep{1}{2} V_{34}
+\cw\ep{3}{4}V_{12}-{2d\over\mw}\cw V_{12}V_{34}\cf\dh{s} \eqa

\bqa & & F_H(ZZ\to ZZ)={g^2_2\mw\over c_W^4}\biggm\{
\ep{1}{2}\ep{3}{4}\co{1\over s-\mw}-\dh{s}\cf+  \nonumber\\
 & &\null +\ep{1}{3}\ep{2}{4}\co
 {1\over t-\mw}-\dh{t}\cf
  +\ep{1}{4}\ep{2}{3}\co
 {1\over u-\mw}-\dh{u}\cf    \biggm\}+ \nonumber\\
 & &\null +2dg^2_2\biggm\{ \co \ep{3}{4} V_{12}
+ \ep{1}{2}V_{34}-{2dc^4_W\over\mw} V_{12}V_{34}\cf\dh{s}+\nonumber\\
 & &\null + \co \ep{2}{4} V_{13}
+ \ep{1}{3}V_{24}-{2dc^4_W\over\mw} V_{13}V_{24}\cf\dh{t}+\nonumber\\
 & &\null + \co \ep{1}{4} V_{23}
+ \ep{2}{3} V_{14}-{2dc^4_W\over\mw} V_{14}V_{23}\cf\dh{u}\biggm\} \eqa

\bqa
 & & F_H(W^+W^-\to W^+W^-)=g^2_2\mw \biggm\{
\ep{1}{2}\ep{3}{4}\co{1\over s-\mw}-\dh{s}\cf+  \nonumber\\
 & &\null +\ep{1}{3}\ep{2}{4}\co
 {1\over t-\mw}-\dh{t}\cf  \biggm\}\nonumber\\
 & &\null +2dg^2_2\biggm\{ \co \ep{3}{4} V_{12}
+ \ep{1}{2}V_{34}-{2d\over\mw} V_{12}V_{34}\cf\dh{s}+\nonumber\\
 & &\null + \co \ep{2}{4} V_{13}
+ \ep{1}{3}V_{24}-{2d\over\mw} V_{13}V_{24}\cf\dh{t} \biggm\} \eqa

\bqa
 & & F_H(W^+W^+\to W^+W^+)=g^2_2\mw \biggm\{
\ep{1}{4}\ep{2}{3}\co{1\over u-\mw}-\dh{u}\cf+  \nonumber\\
 & &\null +\ep{1}{3}\ep{2}{4}\co
 {1\over t-\mw}-\dh{t}\cf  \biggm\}\nonumber\\
 & &\null +2dg^2_2\biggm\{ \co \ep{1}{4} V_{23}
+ \ep{2}{3}V_{14}-{2d\over\mw} V_{14}V_{23}\cf\dh{u}+\nonumber\\
 & &\null + \co \ep{2}{4} V_{13}
+ \ep{1}{3}V_{24}-{2d\over\mw} V_{13}V_{24}\cf\dh{t} \biggm\} \eqa

\bq
F_H(W^+W^-\to\gamma\gamma)={2g^2_2\sw d\over
M_W}V_{34}\co\ep{1}{2}M_W-{2d\over M_W}V_{12}\cf\dh{s}  \eq

\bqa
F_H(W^+W^-\to\gamma Z)&=&{c_W\over s_W}
F_H(W^+W^-\to\gamma\gamma)=\nonumber\\
&=& {2g^2_2s_Wc_W d\over
M_W}V_{34}\co\ep{1}{2}M_W-{2d\over M_W}V_{12}\cf\dh{s} \eqa

\bqa
& &F_H(W^+W^-\to
ZZ)=g^2_2{\mw\over\cw}\ep{1}{2}\ep{3}{4}
\co{1\over s-\mw}-\dh{s}\cf+ \nonumber\\
& &\null
+2g^2_2d\co{1\over\cw}\ep{3}{4}V_{12}+\cw\ep{1}{2}V_{34}-{2d\cw\over\mw}
V_{12}V_{34}\cf\dh{s}   \eqa

\bq F_H(\gamma\gamma\to\gamma\gamma)=
-{4g^2_2s^4_Wd^2\over\mw}\bigg\{ {V_{12}V_{34}\over\nh{s} }
 +{V_{13}V_{24}\over\nh{t}}+{V_{14}V_{23}\over\nh{u} } \biggm\}\eq

\bq F_H(\gamma Z\to\gamma\gamma)={c_W\over s_W}
F_H(\gamma\gamma\to\gamma\gamma) \eq

\bq F_H(ZZ\to\gamma\gamma)=
{2g^2_2s^2_W d\over M_W}\,\bigg\{ {Z_{12}V_{34}\over\nh{s} }
 -{2\cw d\over M_W}\co{V_{13}V_{24}\over\nh{t}}+{V_{14}V_{23}\over\nh{u}}\cf
 \biggm\}\eq

\bq F_H(\gamma \gamma\to\gamma Z)={c_W\over s_W}
F_H(\gamma\gamma\to\gamma\gamma) \eq

\bq F_H(\gamma Z\to\gamma Z)=
{2g^2_2s^2_W d\over M_W}\,\bigg\{ {Z_{24}V_{13}\over\nh{t} }
 -{2\cw d\over M_W}\co{V_{12}V_{34}\over\nh{s}}+{V_{14}V_{23}\over\nh{u}}\cf
 \biggm\}\eq

\bq F_H(Z Z\to\gamma Z)=
{2g^2_2s_Wd\over M_W}\,\bigg\{ {Z_{12}V_{34}\over\nh{s} }
 +{V_{13}Z_{24}\over\nh{t}}+{Z_{14}V_{23}\over\nh{u} } \biggm\}\eq

\bq F_H(\gamma \gamma\to Z Z)=
{2g^2_2s^2_W d\over M_W}\,\bigg\{ {Z_{34}V_{12}\over\nh{s} }
 -{2\cw d\over M_W}\co{V_{13}V_{24}\over\nh{t}}+{V_{14}V_{23}\over\nh{u}}\cf
 \biggm\}\eq

\bq F_H(\gamma Z\to Z Z)=
{2g^2_2 s_Wc_W d\over M_W}\,\bigg\{ {V_{12}Z_{34}\over\nh{s} }
 +{V_{13}Z_{24}\over\nh{t}}+{V_{14}Z_{23}\over\nh{u} } \biggm\}\eq

\newpage

{\Large\bf Figure captions}\\
\begin{description}
\item[{Fig.1}]  Invariant mass distribution in $W^+Z$ production via $W^+Z \to
 W^+Z$.
Standard Model (solid), including $\O_W$ ($\lambda =0.01$ dashed, $\lambda
 =-0.01$
 short
dashed), including $\O_{UW}$ ($d = \pm 0.1$ long dashed).
\item[{Fig.2}] Invariant mass distribution in $ZZ$ production via various
fusion
 processes.
Standard Model(solid), including $\O_W$ ($\lambda =0.01$ dashed, $\lambda
=-0.01$
 short
dashed), including $\O_{UW}$ ($d = \pm 0.1$ long dashed).
\item[{Fig.3}] Invariant mass distribution for various final
states summing over all  possible
initial states. Standard Model (solid), including $\O_W$
($\lambda =0.01$ dashed). The results for $\lambda=-0.01$ are
similar to those for $\lambda=0.01$.
\item[{Fig.4}] Invariant mass distribution for various final states summing all
 possible
initial states. Standard Model (solid), including $\O_{UW}$ ($d = \pm 0.1$ long
 dashed).
\item[{Fig.5}] Ratios of invariant mass distributions $WZ/ \gamma \gamma$ and
 $WZ/ W \gamma$.
Standard Model (solid), including $\O_W$ ($\lambda =0.01$ dashed,
 $\lambda =-0.01$
 short
dashed), including $\O_{UW}$ ($d = \pm 0.1$ long dashed).
\item[{Fig.6}] Ratios of invariant mass distributions $WZ / ZZ$
and $W \gamma /ZZ$.
Standard Model (solid), including $\O_W$ ($\lambda =0.01$ dashed,
 $\lambda =-0.01$
 short
dashed), including $\O_{UW}$ ($d = \pm 0.1$ long dashed).
\end{description}

\end{document}